\documentclass[twocolumn,aps,showpacs,prd]{revtex4-1}
\usepackage{graphicx}
\usepackage{amsbsy}
\usepackage{color}
\usepackage{epsfig}
\usepackage{graphicx}
\usepackage{amsmath}
\usepackage{amssymb}
\usepackage{bbm}
\usepackage{amsbsy}
\usepackage{slashed}

\usepackage{xcolor}

\definecolor{mbcol}{rgb}{1,0,1}

\usepackage[normalem]{ulem}  
\begin{document}

\title{Super-Strong Coupling NJL model in arbitrary space-time dimensions}
\author{Aftab Ahmad$^{1}$, Angelo Mart\'inez$^2$, Alfredo Raya$^2$}
\affiliation{$^1$Department of Physics, Gomal University, 29220, D.I. Khan, K.P.K., Pakistan.\\                             
$^2$ Instituto de F\'{\i}sica y Matem\'aticas,
Universidad Michoacana de
San Nicol\'as de Hidalgo. Edificio C-3, Ciudad Universitaria, Morelia 58040, Michoac\'an, M\'exico.}

\begin{abstract}
We study chiral symmetry breaking in the Nambu--Jona-Lasinio model regularized in proper-time in arbitrary space-time dimensions through an iterative procedure by writing the gap equation in the form of a discrete dynamical system with the coupling constant as the control parameter. Expectedly, we obtain the critical coupling for chiral symmetry breaking when a nontrivial solution bifurcates away from the trivial one and becomes an attractor. By increasing further the value coupling constant, we observe a second bifurcation where the dynamical solution is no longer an attractor, and observation that holds true in all space-time dimensions. In the super-strong coupling regime, the system becomes chaotic. 
\end{abstract}

 
\maketitle

Quantum chromodynamics (QCD)~\cite{Gross:1973id,Politzer:1973fx} is nowadays accepted to be the theory that describes strong interactions among quarks and gluons. It possesses two important and opposite features, {\em asymptotic  freedom}~\cite{Gross:1973id} in the high energy  domain, where quarks move freely at short distances. Perturbative QCD  is the appropriate framework to approach this regime. On the other hand, there exists {\em quark confinement}~\cite{Gribov:1977wm} in the low-energy and large-distances regime.  
In this case, the fundamental degrees of freedom bind in hadronic bound-states (mesons, baryons and exotic), the coupling becomes strong and the theory highly non-linear. A number of non-perturbative frameworks have been developed to study low-energy QCD, among which we find
Lattice QCD~\cite{Wilson:1974sk}, Schwinger-Dyson equations (SDEs)~\cite{Dyson:1949ha,Schwinger:1951ex} (see Refs.~\cite{ROBERTS1994477,MarisIJMPE2003,ALKOFER2001281,CSFischer2006,BINOSI20091,Adnan2012,Eichmann:2016yit} for reviews of SDEs in hadron physics) and other field-theoretical approaches, like the functional renormalization group~\cite{BERGES2002223,PAWLOWSKI20072831,Gies2012},
 as well as effective models. Nambu-Jona-Lasinio (NJL) model~\cite{Nambu:1961tp,Nambu:1961fr} is a favorite one 
to study the properties of  hadrons mostly in connection with chiral symmetry breaking (see, for instance, the reviews in Refs.~\cite{Klevansky:1992qe,Buballa:2003qv}), and it is the general topic of this article. 
The model is non-renormalizable  and thus, there are several schemes in which the regulator and coupling are selected to match static properties of pions (see, for example, Ref.~\cite{Kohyama:2016fif}). Here, we select to regulate the gap equation in the proper-time (PT) scheme. Momentum integrals become Gaussian and thus are readily performed; the regulator comes in the PT integration, as opposed to the more frequently employed  hard momentum cut-off regularization scheme~\cite{Martinez:2018snm}~(the UV regulator in momentum space in the latter case is identified with the IR cut-off in proper-time integrals). Furthermore, by introducing an IR cut-off in momentum  (UV in PT), we mimic confinement in the model ~\cite{Ebert:1996vx,GutierrezGuerrero:2010md,Roberts:2011wy}; the second cut-off removes quark-antiquark production thresholds, in the deep infra-red region of the quark propagator. This procedure has an added advantage that the resulting quark propagator naturally fulfills  the Axial-Vector Ward identities for bound states.
Furthermore, the PT regularization scheme allows an easy incorporation of plasma effects, including finite temperature, density and external magnetic fields. 
Among other things, these extreme conditions make the effective dynamics of quarks reduced from the usual 4-dimensions in vacuum; a strong magnetic field reduces two dimensions, whereas a heat bath at high temperature reduces one dimension. On top of that, the coupling of the original theory effectively changes when quarks interact in a medium
in such a manner that these external agents, temperature and magnetic field, compete with each other to either inhibit
or promote the dynamical breaking of chiral symmetry, namely, the effective coupling
is dressed by the medium in such a way that its strength can be enhanced or diluted.
Therefore, it is natural to ask for the behavior of the solution to the gap equation in several space-time dimensions and for a broad range of values of the coupling~(see, for instance, Refs.~\cite{Klevansky:1992qe, Marquez:2015bca,Ahmad:2016iez,Shovkovy:1995td,Khunjua:2017khh,Zhukovsky:2017hzo,Khunjua:2017mkc,Dudas:2006qv,Antonyan:2006vw}) and explore how scenario of ChSB changes. This is the goal of the present article.

Regarding its parametric structure, NJL model  
has a single coupling constant that must exceed a critical value $G_c$ to describe chiral symmetry breaking. 
That is to say, for $G>G_c$, a non-trivial solution to the quark gap equation $M\ne 0$ bifurcates from the trivial one $M=0$. In the chiral limit (see below), the gap equation can be cast in the form $M=f(M;G,\Lambda,\ldots)$ where the form of $f(M;G,\Lambda,\ldots)$ depends on the regularization scheme through the cut-off(s) $\Lambda$, whereas the coupling constant $G$ plays the role of a control parameter in the language of discrete dynamical systems~\cite{strogatz} and the dots stand for any other possible parameters under consideration, like temperature, density, magnetic fields and so on. A solution $M\ne 0$ turns  out to be an attractor in the sense that regardless the initial condition $M_0$, after a finite number of iterations, the succession $M_0, \ M_1=f(M_0), \ldots,\ M_n=f^{(n)}(M_0)$ is such that $M_n$ lays within a arbitrarily small vicinity around the solution $M$. Thus, this procedure is useful to find the solution to the gap equation with a desired precision merely by considering more and more iterations.
In Ref.~\cite{Aoki:2013gda}, for example, by using a hard momentum cut-off regularization scheme, the authors cast the gap equation for the NJL model in terms of a discrete dynamical system with the coupling of the model as control parameter and interpret each iteration as a self-energy insertion in the quark propagator. They  identify the critical value of the coupling for chiral symmetry breaking, $G_c$, consistent with other approaches and find the dynamically generated mass to be an attractor even for large values of the coupling constant $G>G_c$.   These findings were generalized by some of us in Ref.~\cite{Martinez:2017wdu} 
for different regularization schemes. In that work, we observed, not surprisingly, that for the three-dimensional (3D) and four-dimensional (4D) hard momentum cut-off regularization schemes, the procedure converges to the dynamical mass in a finite number of iterations and this solution continues to be an attractor even in  the super-strong coupling regime. On the other hand,  for the Pauli-Villars (PV) and PT regularization schemes, we noticed that a second bifurcation appears for a coupling larger than $G_c$ such that the solution ceases to be and attractor and a double-period orbit develops. For even larger values of the coupling, the system becomes chaotic. 
This means that due to the non-renormalizability of the model, care has to be taken when resumming self-energy corrections and is the point we pursue to stress in the manuscript: When high-momentum modes
in the self-energy are not removed to start with, the resummation is no longer stable.
Our goal in this article thus becomes to explore in a closer look the circumstances under which the first and second bifurcations of the solutions to the  gap equation in  the super-strong coupling regime take place in arbitrary space-time dimensions. 
For that purpose, we have organized the rest of the manuscript as follows: In Sect.~\ref{sec:NJL} we present the  basics  of the  NJL model in vacuum for arbitrary space-time dimensions. The general iterative procedure to solve the gap equation is discussed in Sect.~\ref{sec:sol}. Numerical findings of the evolution of the first and second bifurcations in arbitrary space-time dimensions are discussed too. 
The analysis for the confining variant of the model is discussed in Sect.~\ref{secCI}.
We present a summary of our findings and perspectives in Sect.~\ref{conclusions}.

\section{Nambu--Jona-Lasinio Model}\label{sec:NJL}

\begin{figure}[t!]
\begin{center}
\includegraphics[width=0.9\columnwidth]{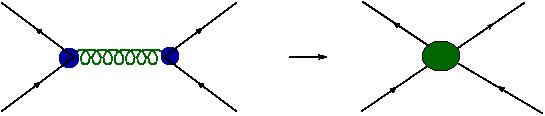}
\caption{Schematic representation of one-gluon exchange  replaced by a four-fermion contact interaction in the NJL model.}
\label{figNJL1}
\end{center}
\end{figure}

\begin{figure}[t!]
\begin{center}
\includegraphics[width=0.9\columnwidth]{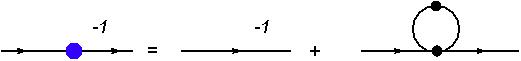}
\caption{The pictorial representation of the gap equation for the NJL model.}
\label{figNJL2}
\end{center}
\end{figure}

The Nambu--Jona-Lasinio (NJL) model~\cite{Nambu:1961tp,Nambu:1961fr} was first introduced to explain chiral symmetry breaking  for nucleons inspired by the phenomenon of spontaneous symmetry breaking in superconductors even before the QCD era. In modern considerations, it is now regarded as an effective model of interactions among quarks and gluons in which the  one gluon exchange between the quarks is replaced by  four-fermion contact interaction, as shown in pictorial form in  Fig.~\ref{NJL1}.  
The Lagrangian of the NJL model for two light quark flavors $N_f=2$, and  three colors $N_c=3$, is given by~\cite{Klevansky:1992qe,Buballa:2003qv}
\begin{equation}
{\cal L}=\bar\psi (i\slashed {\partial}-m)\psi+ \frac{G}{2}[(\bar\psi\psi)^2+(\bar\psi i\gamma_5 \vec{\tau}\psi)^2],\label{NJL1}
\end{equation}
where the first term is the matter (Dirac) part of the Lagrangian with $ m={\rm diag}(m_u, m_d)$ denoting the current light quark mass, which in the chiral limit we set $m=0$. The second term is the four-fermion interaction  
with
$(\bar\psi\psi)^2$ is the  scalar piece of the  interaction, while  $(\bar\psi i\gamma_5 \vec{\tau}\psi)^2$ is  axial-vector  part. Here  $\vec{\tau}$ are the Pauli  matrices acting in  isospin space and $G$ denotes the  coupling constant. The above Lagrangian~(\ref{NJL1}) respects all the  global symmetries of the the full QCD Lagrangian in chiral limit~\cite{Klevansky:1992qe,Buballa:2003qv}. Chiral symmetry can be explicitly broken via a finite current quark mass. It can also be broken spontaneously. In this regard, a starting point to study the latter scenario is the gap equation, shown pictorially in Fig.~\ref{figNJL2}.

From a Schwinger-Dyson equations (SDEs) point of view, we start from the most general form of the dressed quark propagator
\begin{eqnarray} 
S^{-1}(p)=i\gamma \cdot p +m+\Sigma(p),\label{NJL6} 
\end{eqnarray}
where  $\Sigma(p)$ represents the quark self-energy, which in QCD corresponds to the expression
\begin{eqnarray}
\Sigma(p)=\int_\Lambda{\frac{d^D k}{(2\pi)^D }} g^2 D_{\mu\nu} ((k-p)^2 ){\frac{\lambda^a}{2}} \gamma_\mu S(k){\frac{\Lambda^a}{2}}\Gamma_\nu (k,p).\nonumber\\
\label{NJL7} 
\end{eqnarray}
with  $D_{\mu\nu} (q^2)$ denoting the gluon propagator, $g^2$ the QCD coupling constant, $\Gamma_\nu (k,p)$ is the  dressed quark-gluon  vertex,  $\lambda^{a}$ are the $SU(3)$  Gell-Mann matrices. Mapping one-gluon exchange diagrams to a contact interaction 
in the NJL model equivalent to consider
\begin{eqnarray}
g^2 D_{\mu\nu} (q^2) \rightarrow G \delta_{\mu\nu} \Theta(k^2-\Lambda^2), \qquad \Gamma_\nu (k,p)=\gamma_\nu,
 \label{NJL8} 
\end{eqnarray}
in Eq.~(\ref{NJL7}). Then, it is straightforward to verify that there are no wavefunction renormalization effects in this model, and that the quark mass function is merely a constant such that the dressed quark propagator takes the form
\begin{eqnarray}
S(k) = \frac{i\gamma \cdot k +M} {k^2 + M^2},\label{NJL9} 
\end{eqnarray}
with $M$  denoting the momentum independent dynamical quark mass, which is self-consistently determined from the gap equation
\begin{eqnarray}
M&=&m+2G N_f N_c \int_{\Lambda}
\frac{d^Dk}{(2\pi)^D}  {\rm Tr_{D}[S(k)]}\nonumber\\
&=&m+2DG N_f N_c \int_{\Lambda}
\frac{d^Dk}{(2\pi)^D} \frac{M}{k^2+M^2}
. \label{NJL10}
\end{eqnarray}
The symbol $\int_{\Lambda}$ stresses that the model is not renormalizable and we need to regulate the integrals involved.
Within the PT regularization scheme, 
we exponentiate the denominator of the integrand in~(\ref{NJL10}) such that  
\begin{eqnarray}
\int_{\Lambda} \frac{d^D k}{(2\pi)^D}\frac{1}{k^2+M^2}&\to&\int\frac{d^D k}{(2\pi)^D}\int^{\infty }_{\Lambda^2} d\tau {\rm e}^{-\tau(k^2+M^2)}\nonumber\\
&=&\frac{1}{(4\pi)^\frac{D}{2} }   \int^{\infty }_{\Lambda^2}\frac{ d\tau}{\tau^{D/2}} {\rm e}^{-\tau M^2} \label{NJL13},
\end{eqnarray}
where in the last step we have performed the Gaussian momentum integrations. Notice that $\Lambda$ and $\lambda$  are regulators with mass dimension -1. Thus, the gap equation in arbitrary dimensions can be written as
\begin{eqnarray}
M-m=
G_D M {\rm E_\frac{D}{2}\left(M^2\Lambda^{2}\right)}, \label{NJL14} 
\end{eqnarray}
with $E_n(x)$ is defined as~\cite{abramowitz+stegun}
\begin{equation}
E_n(x)=\int_1^\infty dt \ \frac{e^{-x t}}{t^n}\;,
\end{equation}
and the effective coupling
\begin{eqnarray}
G_D=\frac{DG N_f N_c \Lambda^{2-D}}{(2)^{D-1}(\pi)^{D/2}}.\label{coupD} 
\end{eqnarray}
With these ingredients, and for the sake of exploring the parameter dependence of the gap equation, we rewrite Eq.~(\ref{NJL14})
in dimensionless form as
\begin{eqnarray}
\mu-\mu_0= G_D \mu f(\mu)  \label{NJL15}, 
\end{eqnarray}
with
 $\mu=M \Lambda$, $ 
\mu_0=m\Lambda$ and $f(\mu)= {\rm E_\frac{D}{2}(\mu^2)}$. 

Before proceeding to analyze Eq.~(\ref{NJL15}), let us consider a variant of the regularization in Eq.~(\ref{NJL13}) that has been recently introduced and has the form~\cite{Ebert:1996vx,GutierrezGuerrero:2010md,Roberts:2011wy}
\begin{eqnarray}
\int_\Lambda\frac{d^D k}{(2\pi)^D}\frac{1}{k^2+M^2}&\to&\int\frac{d^D k}{(2\pi)^D}\int^{\lambda^2 }_{\Lambda^2} d\tau {\rm e}^{-\tau(k^2+M^2)}\nonumber\\
&=&\frac{1}{(4\pi)^\frac{D}{2} }   \int^{\lambda^2 }_{\Lambda^2}\frac{ d\tau}{\tau^{D/2}} {\rm e}^{-\tau M^2} \label{double}.
\end{eqnarray}
This double cut-off PT regularization has the advantage of mimicking confinement. Unlike the standard regularization in Eq.~(\ref{NJL13}), the form of the quark propagator regularized as in Eq.~(\ref{double}) is consistent with the Axial-Vector-Ward identities when used in the Bethe-Salpeter equation~\cite{GutierrezGuerrero:2010md,Roberts:2011wy}. This model is dubbed {\em Contact Interaction (CI)} model, and its coupling in four-dimensions is parameterized as $G^{{\rm CI}}=4 \pi \alpha_{ir}/m^{2}_{G}$,
where $m_G = 800 \ {\rm MeV} $ is a  dynamical gluon mass parameter~\cite{Bowman:2004jm} and  
$\alpha_{ir} = 0.93 \pi$ is the interaction strength. For this CI model, the gap equation reads
\begin{eqnarray}
M=m+\frac{ G^{\rm CI}D}{(4\pi)^\frac{D}{2} }   \int^{\lambda^2 }_{\Lambda^2}\frac{ d\tau}{\tau^{D/2}} {\rm e}^{-\tau M^2}, \label{CI4a}
\end{eqnarray}
which in terms of the dimensionless variables $\mu$ and $\mu_0$ described above and defining $\rho=\lambda / \Lambda$, we re-write  as
\begin{eqnarray}
\mu-\mu_0&=&  G^{\rm CI}_{D}\mu\left({\rm E_\frac{D}{2}(\mu^2)}- \rho^{2-D}{\rm E_\frac{D}{2}(\mu^2\rho^{2})} \right)\nonumber\\
&\equiv& G^{\rm CI}_{D}\mu f^{{\rm CI}}(\mu),\label{CI5}
\end{eqnarray}
with $G^{{\rm CI}}_{D}=DG^{\rm CI}/(2)^{D-1}(\pi)^{D/2}$. In the following, we study the solutions of Eqs.~(\ref{NJL15}) and~(\ref{CI5})  through an iterative procedure.

\section{Solution of the gap equation}\label{sec:sol}
\subsection{Analysis of NJL model in $D=4$}

\begin{figure}[t!]
\begin{center}
\includegraphics[width=0.9\columnwidth]{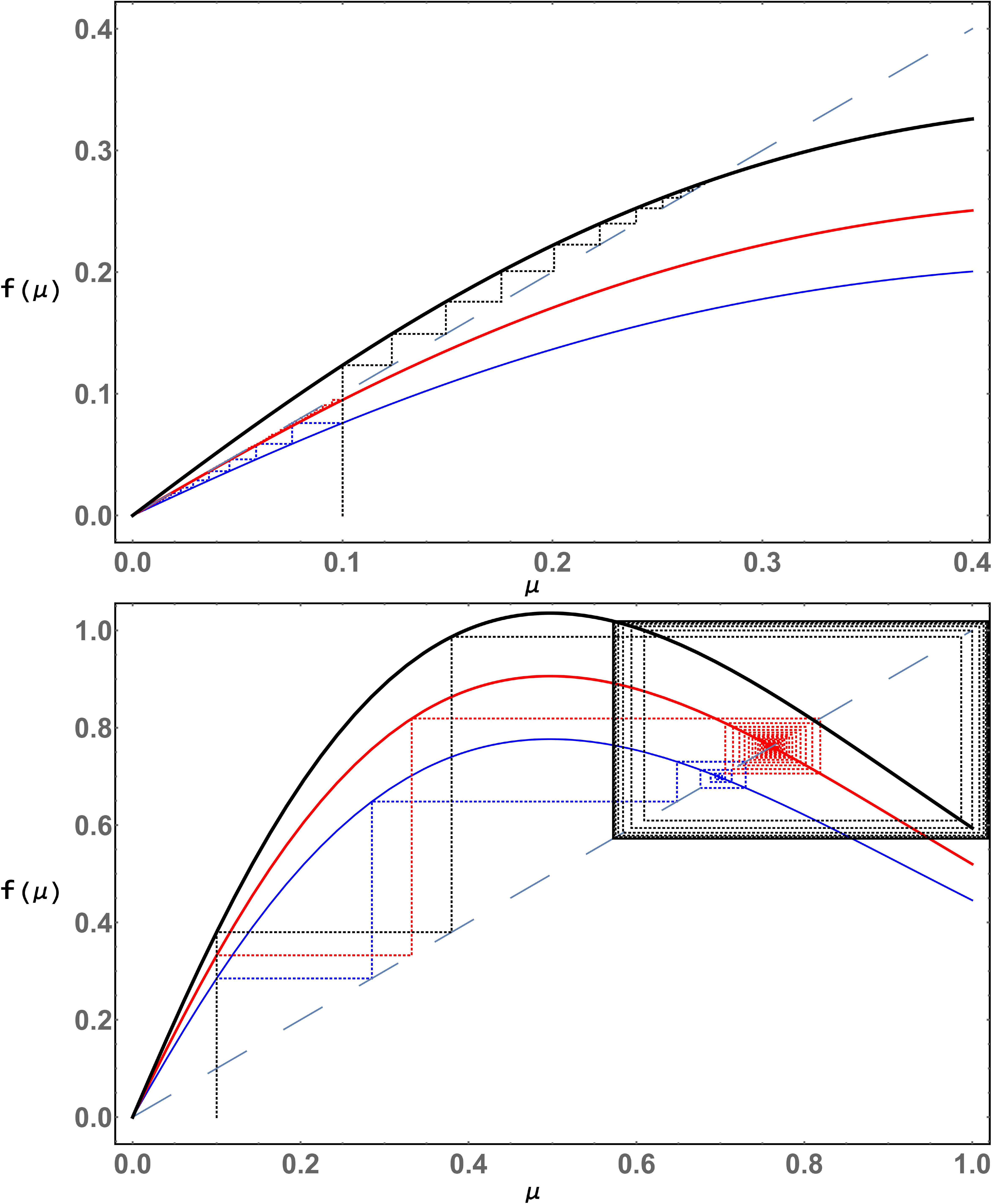}
\caption{Cobweb plot of Eq.~(\ref{NJL15}) in the chiral limit $\mu_0=0$. {\em Upper panel:} Blue (thin) solid line corresponds to the r.h.s. of the gap eq.~(\ref{NJL15}) as a function of $\mu$ at fixed $G_4=0.8$, red solid line to the  $G_4=1$ and black (thick) line to $G_4=1.3$. The diagonal line $y=\mu$ (dashed) is also plotted. Here, the solution of the gap equation are represented by the intersections of the corresponding curves with the dashed line. {\em Lower panel:}  Blue (thin) solid line corresponds to $G_4=3$, red solid line to $G_4=3.5$ and black (thick) line to $G_4=4$. Here, it is obvious
that the sequence of iterations tends to converge to the intersection of the curves until a 2-cycle is reached at.
}
\label{figNJS}
\end{center}
\end{figure}

\begin{figure}[t!]
\begin{center}
\includegraphics[width=0.9\columnwidth]{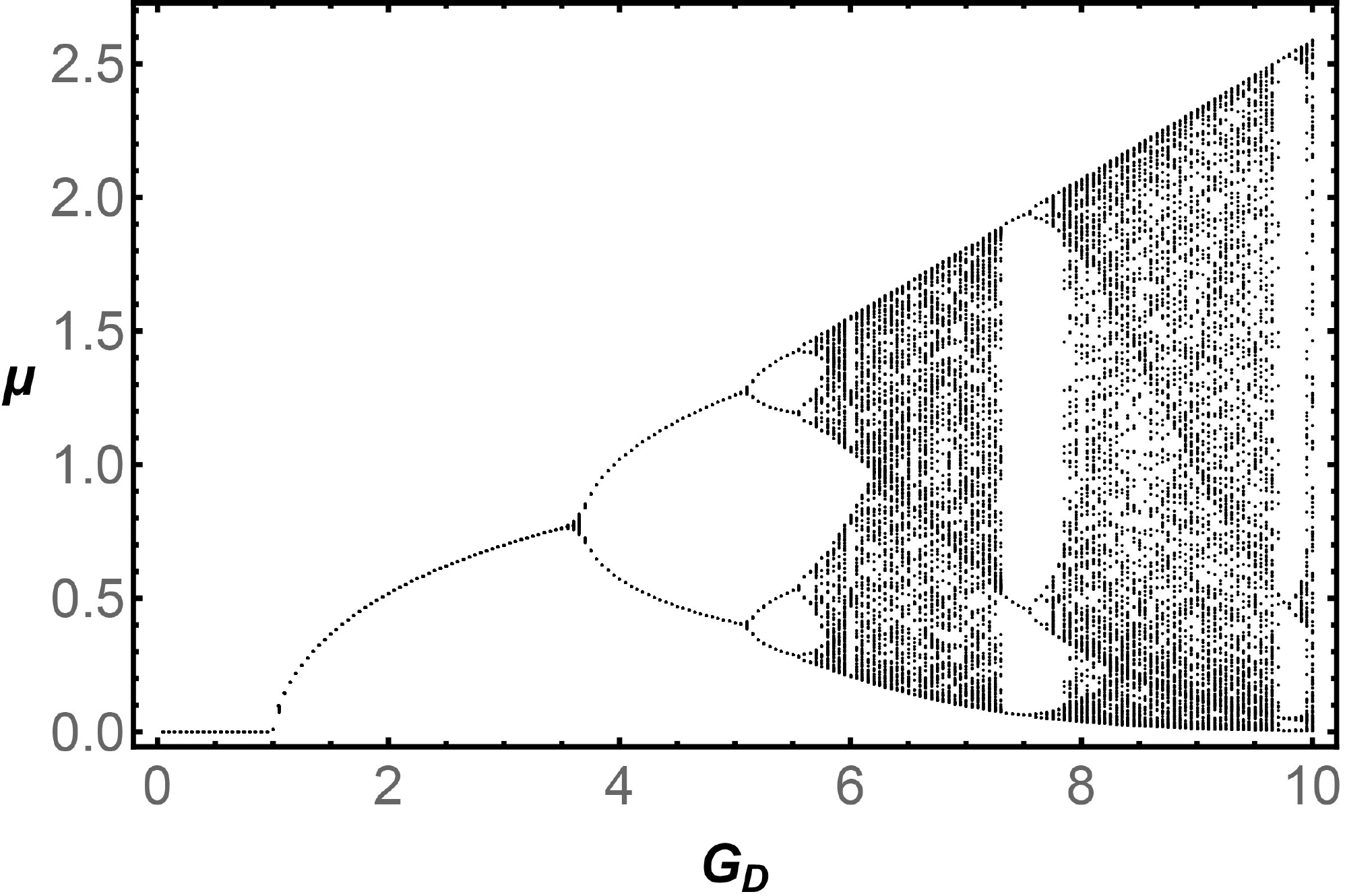}
\caption{Bifurcation map corresponding to the gap eq.~(\ref{NJL15}) in the chiral limit. At fixed $G_4$, given an arbitrary initial guess $\mu_{\rm ini}$, we perform 200 iterations of the mapping (\ref{NJL15}) in the chiral limit and plot only the final 50 iterations. For $G_4<1$, the iterative procedure converges to the value $\mu=0$. For $1<G_4<3.5$ there is a stable solution $\mu\ne 0$ which corresponds to a chiral symmetry breaking solution and has a fixed-point nature. For $G_4>3.5$ a cascade of period-doubling orbits appears, till the system becomes chaotic.}
\label{figNJD4}
\end{center}
\end{figure}

\begin{figure}[t!]
\begin{center}
\includegraphics[width=0.9\columnwidth]{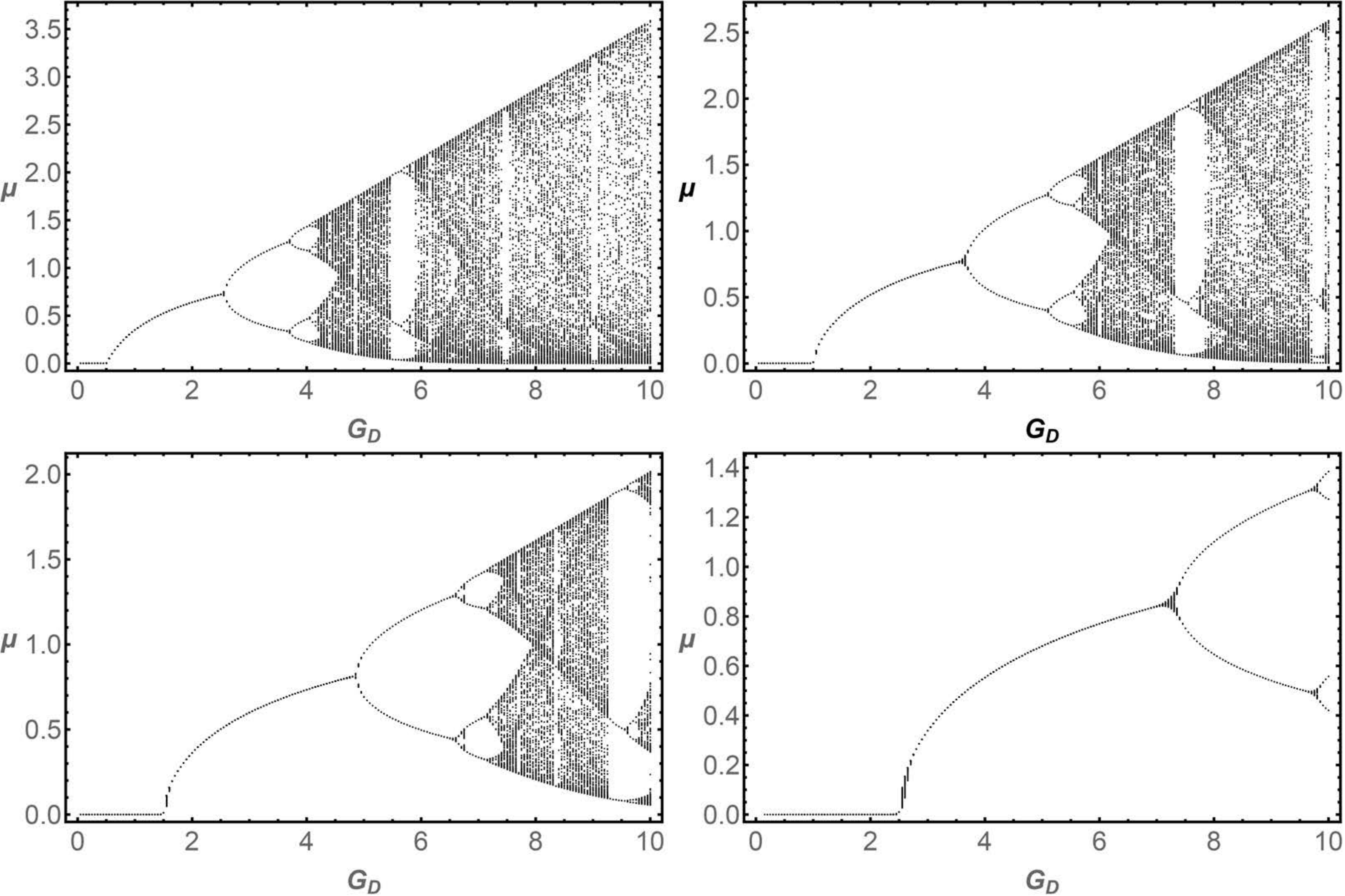}
\caption{Bifurcation map corresponding to the gap eq.~(\ref{NJL15}) in the chiral limit in different dimensions. {\em Upper left panel:} $D=3$; {\em upper right panel:} $D=4$; {\em lower left panel:} $D=5$; {\em lower right panel:} $D=7$. Procedure is the same as in Fig.~\ref{figNJD4}.
}
\label{figNJs2}
\end{center}
\end{figure}

In the following, and for the sake of illustration, we consider the solutions to the gap Eq.~(\ref{NJL15}) in the chiral limit, $\mu_0=0$,  in the case $D=4$. The ideas presented here shall be extended to other cases below. Let us commence, inspired by the analysis of the logistic map~\cite{strogatz}, by observing that the strength of the coupling, $G_4$ in our notation, dictates the number of solutions and its nature in the dynamical systems sense. In Fig.~\ref{figNJS} we plot  the r.h.s. of the gap equation as a function of $\mu$ for three values of the coupling in Eq.~(\ref{coupD}). The upper panel corresponds to  $G_4=0.8$, $G_4=1$ and $G_4=1.3$. We also plot the curve $y=\mu$ as the l.h.s. of the gap equation and look for the intersection of both these curves at fixed $G_4$. We observe that for $G_4<1$, the only intersection on the corresponding two curves is at the origin $\mu=0$ and the r.h.s curve always lies below the diagonal line. By observing the cobweb plot (also shown), we notice that the trivial solution corresponds to a fixed point. It is an attractor in the sense that given an arbitrary initial guess $\mu_i$, in a finite number of iterations, we can find the iterations in a vicinity of the origin. The r.h.s. curve is taller as $G_4$ grows bigger, being tangent to the diagonal line when $G_4=1$ near the origin. As the coupling increases, the r.h.s. curve intersects in two points, the origin and a new point $\mu^f\ne0$ which turns out to be fixed and an attractor, as seen in the cobweb construction. By still demanding the coupling to be larger (lower panel in  \ref{figNJS} we show the cases $G_4=3$, $G_4=3.5$ and $G_4=4$), the point $\mu^f$ ceases to be fixed at the point $\mu^*$ when the slope of the r.h.s. curve reaches its critical value $f'(\mu^s)=-1$, which turns out to be the case for $G_4=3.5$, resulting in a flip bifurcation~\cite{strogatz}. This is a similar test performed in the logistic map to describe precisely this type of bifurcation. For larger couplings ($G_4=4$ in Fig.~\ref{figNJS}) the cobweb plots shows a 2-cycle orbit, observed as the corners of the black square. A doubling-period cascade develops for larger values of $G_4$ until the system becomes chaotic, as observed in Fig.\ref{figNJD4}.
This picture is observed in arbitrary dimensions (see Fig.~\ref{figNJs2}) and also with the CI model (see Sec.~\ref{secCI}) for different values of the second cut-off. 
Nevertheless, we would like to emphasize that the bifurcation map of Fig.~\ref{figNJD4}  does not correspond to the actual solution of the gap equation. As mentioned before, the procedure of iterating the gap equation is equivalent to the loop expansion of the propagator and with each iteration, the number of loops added to the propagator increases by 1. Following this idea, at one hand, the
convergence of the iterations for relatively small couplings means that higher loops do not contribute and thus the iterative procedure converges to the actual solution to the gap equation in this case, the plot corresponds to the solution of the gap equation. In Fig.~\ref{figNJD4}, it is shown as a stable branch before the flip bifurcation develops. On the other hand, when the coupling increases, the non convergence of the iterations points out that every added loop contributes and thus we got the bifurcation diagram
even though for the gap equation itself there is only one solution, which, by the way, never exceeds the natural scale of the model, namely, the cut-off. 
Below we present a detailed analysis in these cases.

\subsection{Analysis in arbitrary dimensions}

\begin{figure}[t!]
\begin{center}
\includegraphics[width=0.9\columnwidth]{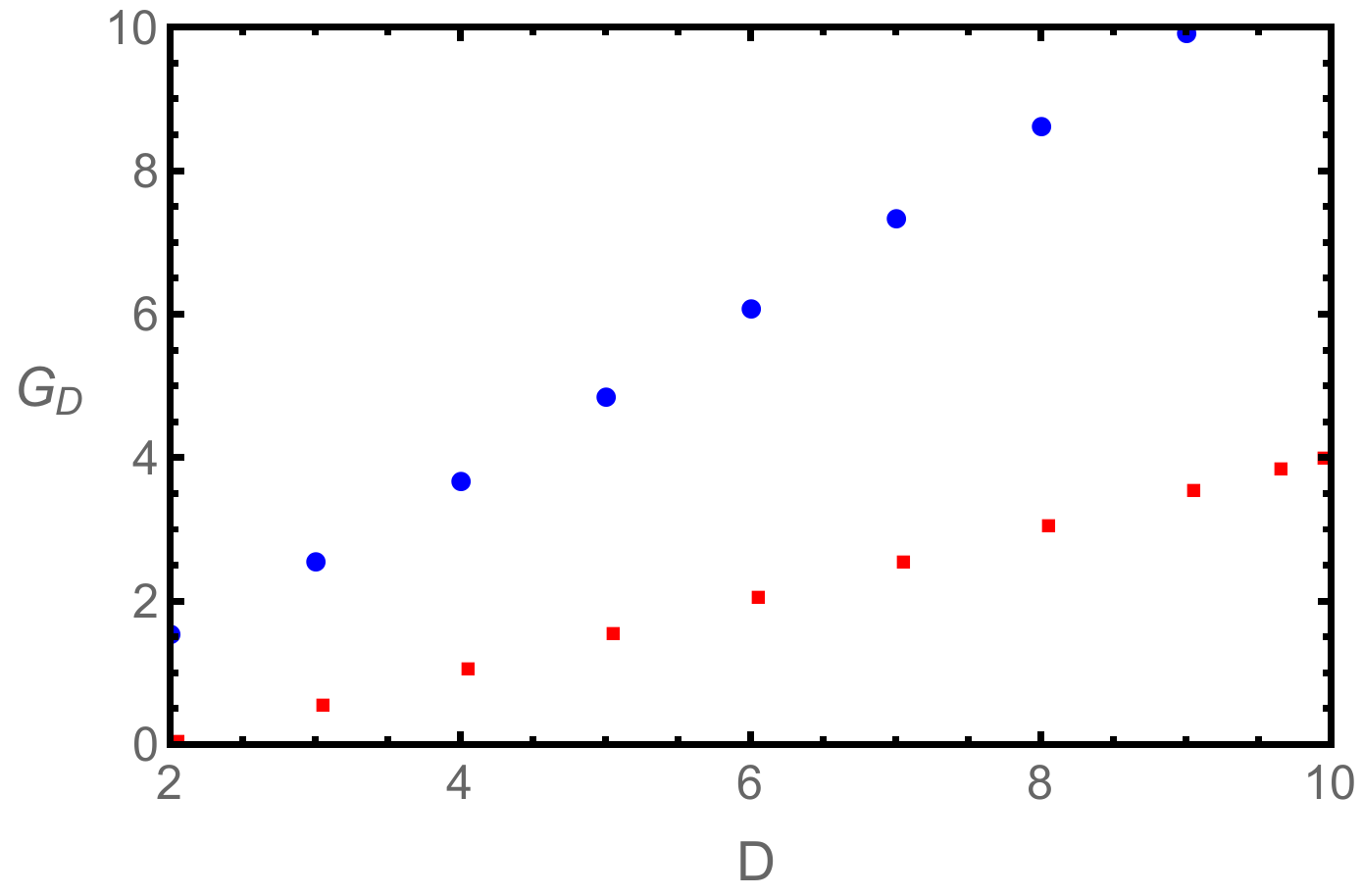}
\caption{Variation of the critical couplings for the first and second bifurcations with $D$:  Squares (red) represent $G^{0}_D$ and  Circles (blue) $G^{*}_{D}$. }
\label{figNJ2}
\end{center}
\end{figure}

The bifurcation map described above replicates in arbitrary dimensions, as can be observed in Fig.~\ref{figNJs2}. Therefore, an analysis similar to the logistic map can be done to identify the first and second bifurcations in the map.
To this end, we start from the gap eq.~(\ref{NJL15}) and take the derivative  with respect to $\mu$ in both sides, namely
\begin{eqnarray}
1= G_D\left( f(\mu)+\mu f^{\prime}(\mu)\right).   \label{NJL17} 
\end{eqnarray}
The critical coupling for the first bifurcation hence appears when we evaluate the above expression at $\mu=0$, when the nontrivial solution bifurcates from the trivial one. Thus, the critical coupling for chiral symmetry breaking in arbitrary dimensions is given as
\begin{eqnarray}
G^{0}_{D}=\frac{1}{ f(0)}  \label{NJL16},
\end{eqnarray}
and it is plotted in Fig.~\ref{figNJ2}. We observe that the slope of the curve of this critical coupling as $D$ increases is uniform, suggesting a linear dependence with $D$.

\begin{figure}[t!]
\begin{center}
\includegraphics[width=0.9\columnwidth]{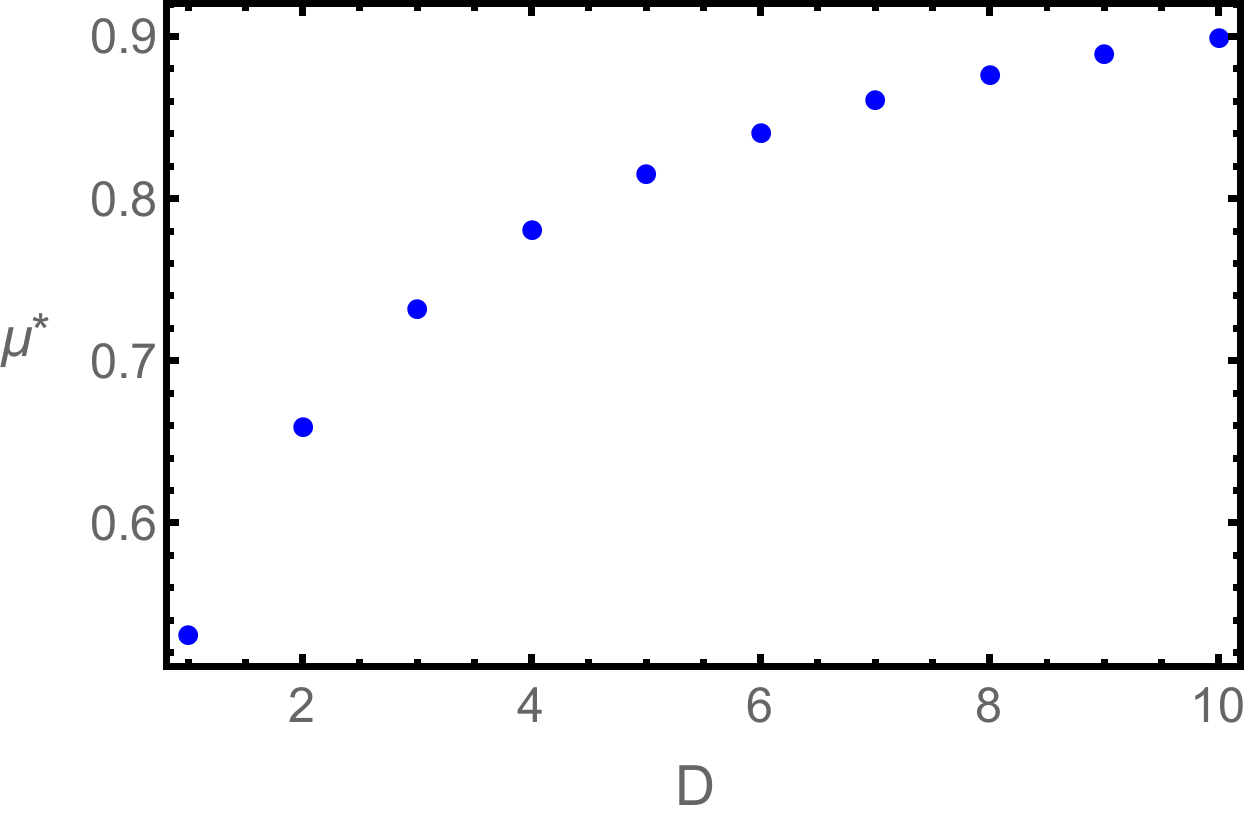}
\caption{Variation of $\mu^{*}$ with $D$. Its growth is faster when $D$ is small and tends to saturate at large $D$.}
\label{figNJ3}
\end{center}
\end{figure}

A standard criterion to locate flip bifurcations in dynamical systems such as the logistic map is found when the slope of the map under study reaches the critical value -1~\cite{strogatz}. We adapt this criterion which in our case consists in locating the position $\mu^*$ where the function 
\begin{eqnarray}
A_D(\mu^*)=\frac{\mu^* f^{\prime}(\mu^*)}{ f(\mu^*)}=-2.   \label{NJL18} 
\end{eqnarray}
The variation of $\mu^{*}$ in different dimensions is shown in Fig.~\ref{figNJ3}. It monotonically increases with $D$ till it saturates at large number of dimensions. The value of the critical coupling for the second bifurcation, $G^{*}_D$,   is 
\begin{eqnarray}
 G^{*}_{D}=\frac{1}{ f(\mu^{*})} \label{NJL19}. 
\end{eqnarray} 
We plot this coupling in Fig.~\ref{figNJ2}. It linearly rises with the dimension $D$, but changes slope around $D\simeq 4$. For dimensions $D$ lower that 4, the stable branch, namely the portion of the plot where the iterative procedure converges to the actual solution of the gap equation before the second (flip) bifurcation appears, is shorter than for $D>4$, as can be seen from the separation between the two couplings.  This means that the larger the dimension, the fixed point branch associated with chiral symmetry breaking is more stable.


\section{Confining model}~\label{secCI}

\begin{figure}[t!]
\begin{center}
\includegraphics[width=0.9\columnwidth]{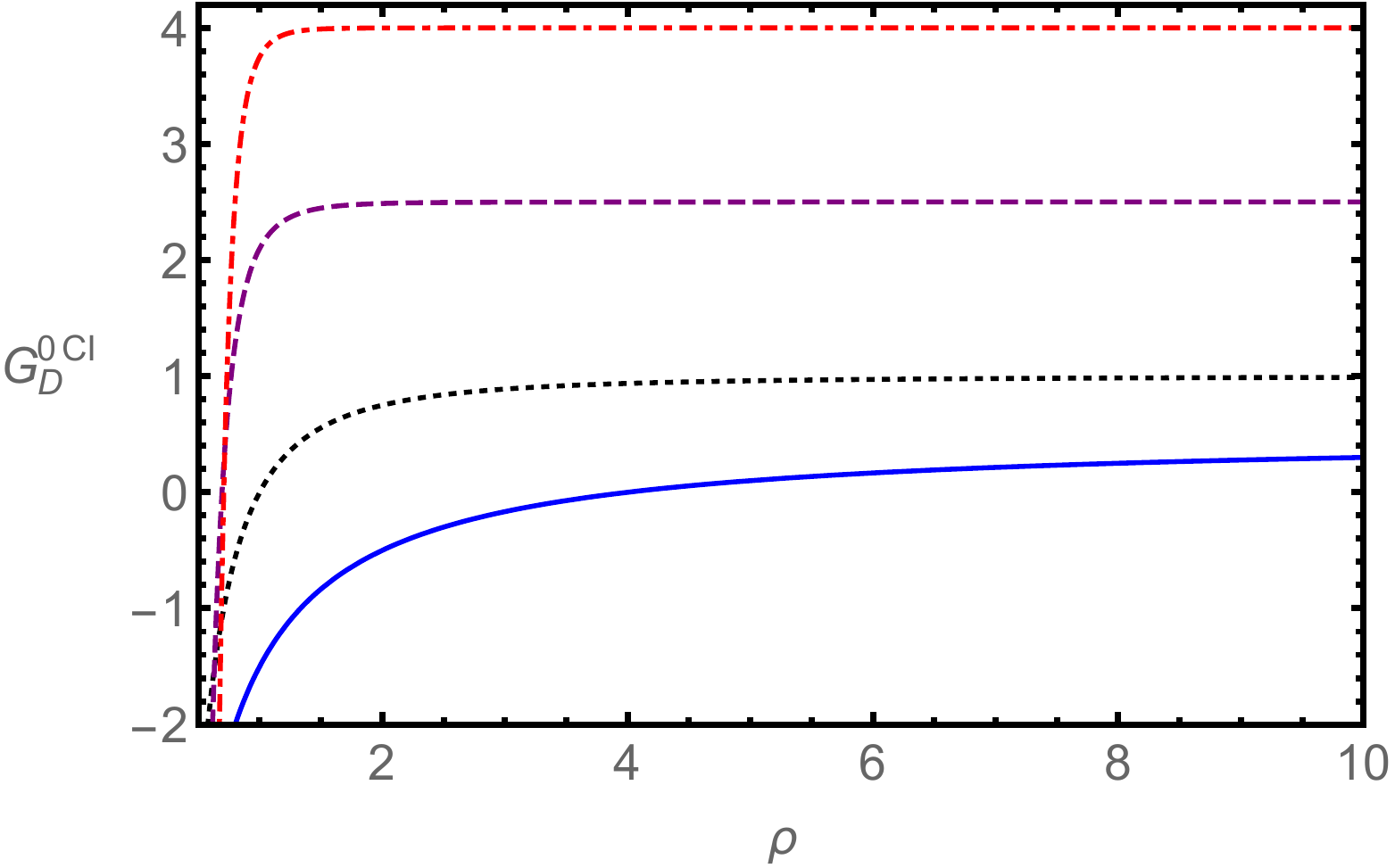}
\caption{$G^{0\ {\rm CI}}_{D}$  as a function of the cut-off $\rho$ for different dimensions $D$}: Solid (blue) curve corresponds to $D=3$; dotted (black) curve to $D=4$; dashed (purple) curve to $D=7$ and dot-dashed (red) curve to $D=10$. 
\label{figCI1}
\end{center}
\end{figure}

\begin{figure}[t!]
\begin{center}
\includegraphics[width=0.9\columnwidth]{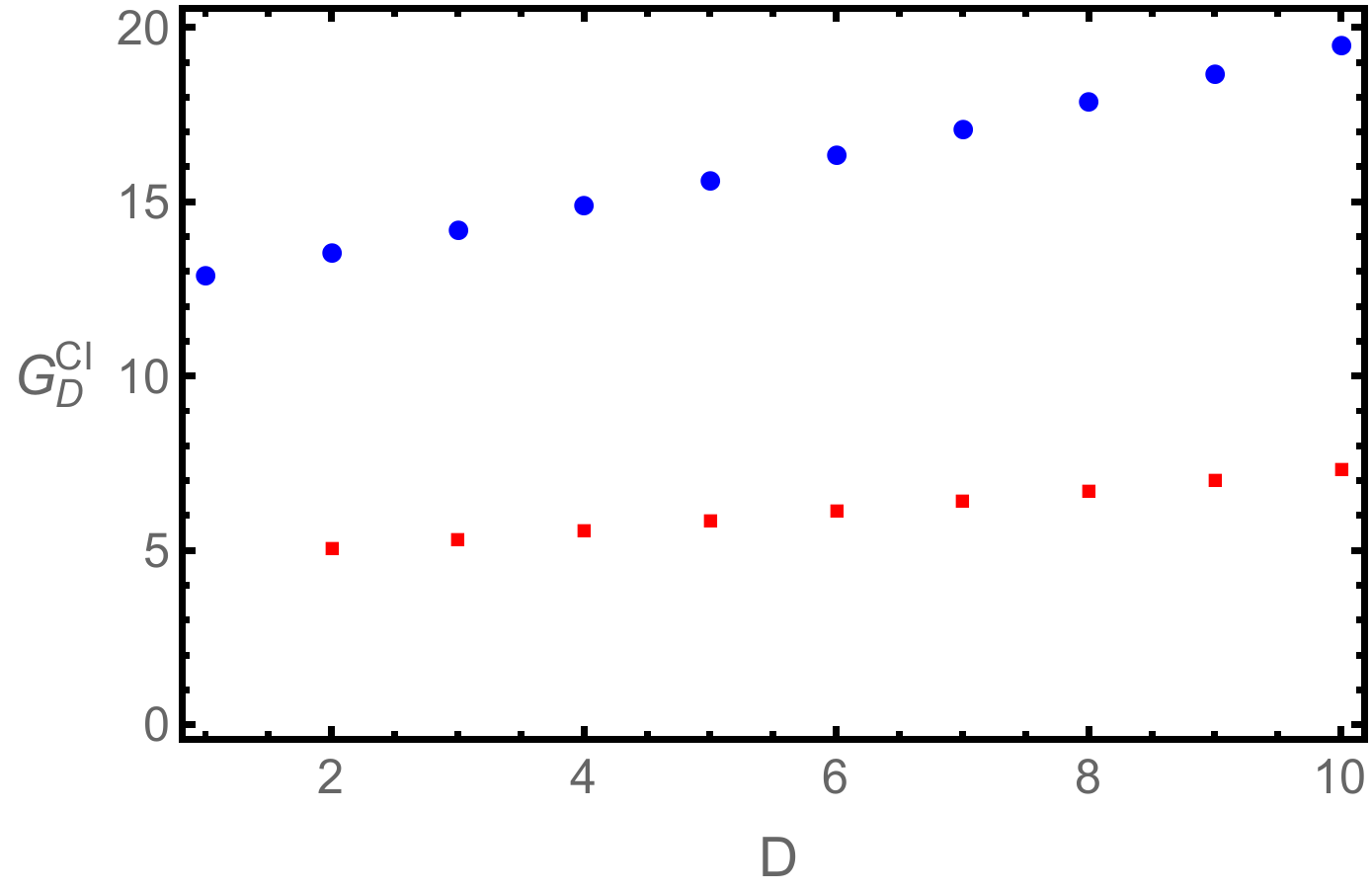}
\caption{Variation of  $G^{CI}_{D}$ with the dimension $D$: Squares (red) represent $G^{0~CI}_{D}$  and circles (blue) $G^{*~CI}_{D}$. We have fixed  $\rho=1.1$.}
\label{figCI2}
\end{center}
\end{figure}

We can extend the above analysis to the confining version of the model described in Eq.~(\ref{CI5}) in the chiral limit $\mu_0=0$. In this case, the derivative of the gap equation should be taken for $f^{\rm CI}(\mu)$. First, we notice as the cut-off $\rho$ grows, the critical value of the coupling for the first branching, namely
\begin{equation}
G_D^{0\ {\rm CI}}=\frac{1}{f^{\rm CI}(0)},
\end{equation}
saturates at a fixed $D$ for its value as in the ordinary model, eq.~(\ref{NJL15}). This is shown in Fig.~\ref{figCI1}, where we observe that $G_D^{0\ {\rm CI}} \to G_D^{0}$ as $\rho\to\infty$. At fixed $\rho$, the variation of $G_D^{0\ {\rm CI}}$ with the dimension is shown in Fig.~\ref{figCI2}. A linear growth is evident to the eye.

\begin{figure}[t!]
\begin{center}
\includegraphics[width=0.9\columnwidth]{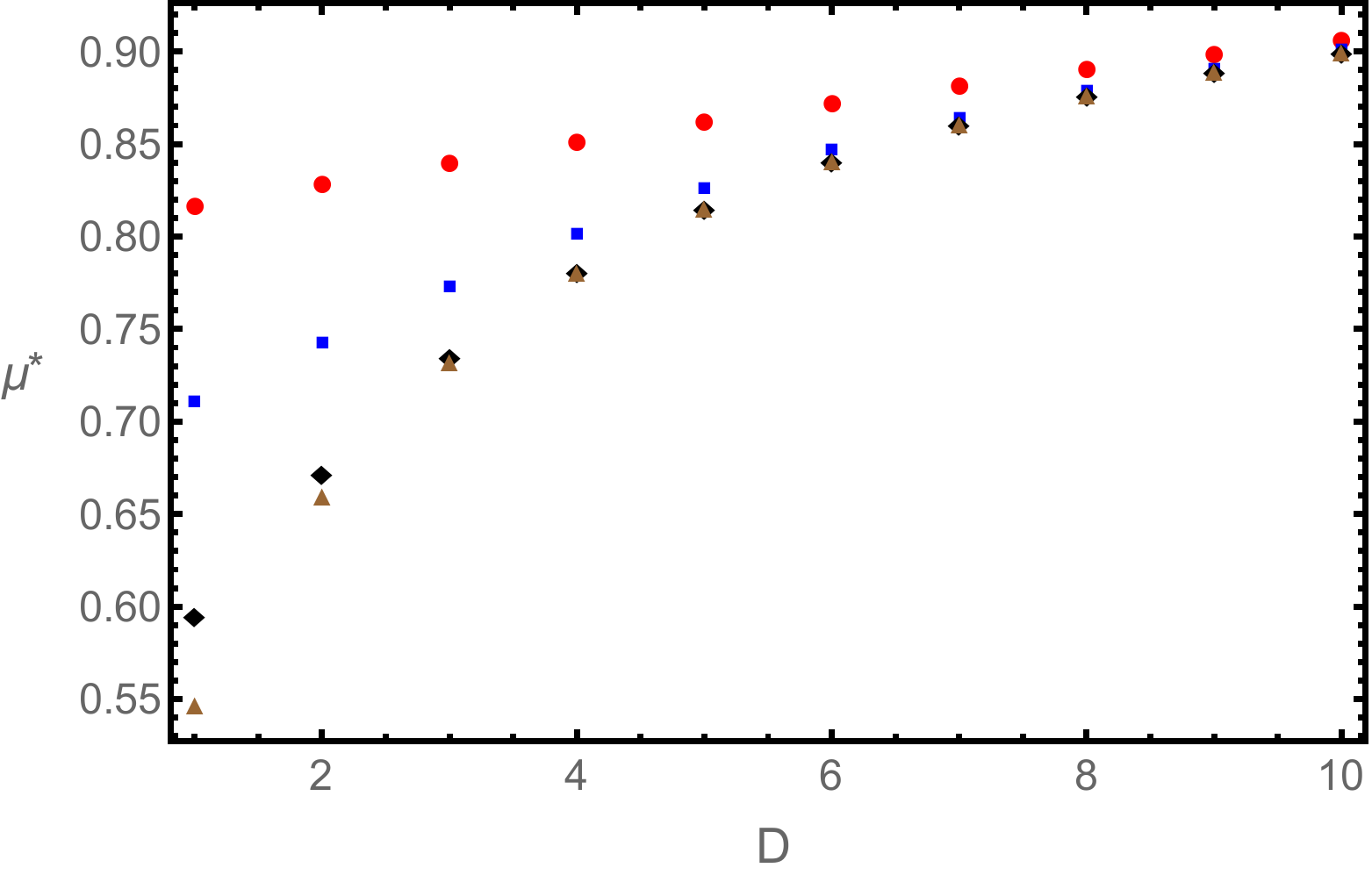}
\caption{Variation of $\mu^{*}_{CI}$ with the dimension $D$: Circles (red) $\rho=1.5$; squares (blue) $\rho=2$; diamonds (black) $\rho=3$; triangles (brown)  $\rho=4$. }
\label{figCI3}
\end{center}
\end{figure}

The critical coupling for the second bifurcation, $G_D^{*\ {\rm CI}}$ is, as before, found by locating the value of $\mu^*$ at which
\begin{equation}
\mu^* \frac{(f^{\rm CI})'(\mu^*)}{f^{CI}(\mu^*)}=-2,
\end{equation} 
in the form
\begin{equation}
G_D^{*\ {\rm CI}}=\frac{1}{f^{\rm CI}(\mu^*)},
\end{equation}
and is also shown in Fig.~\ref{figCI1}. Again, it grows linearly in two regimes separated around $D=4$, where the slope changes such that the first branch of fixed-point solutions is stable for a larger interval of values of $G_D^{\rm CI}$. The scale $\mu^*$ is shown as a function of $D$ in Fig.~\ref{figCI3} for fixed $\rho$. We notice that the impact of the cut-off $\rho$ is less important for large $D$, and that the variation of $\mu^*$ is less for a smaller cut-off.

\begin{figure}[t!]
\begin{center}
\includegraphics[width=0.9\columnwidth]{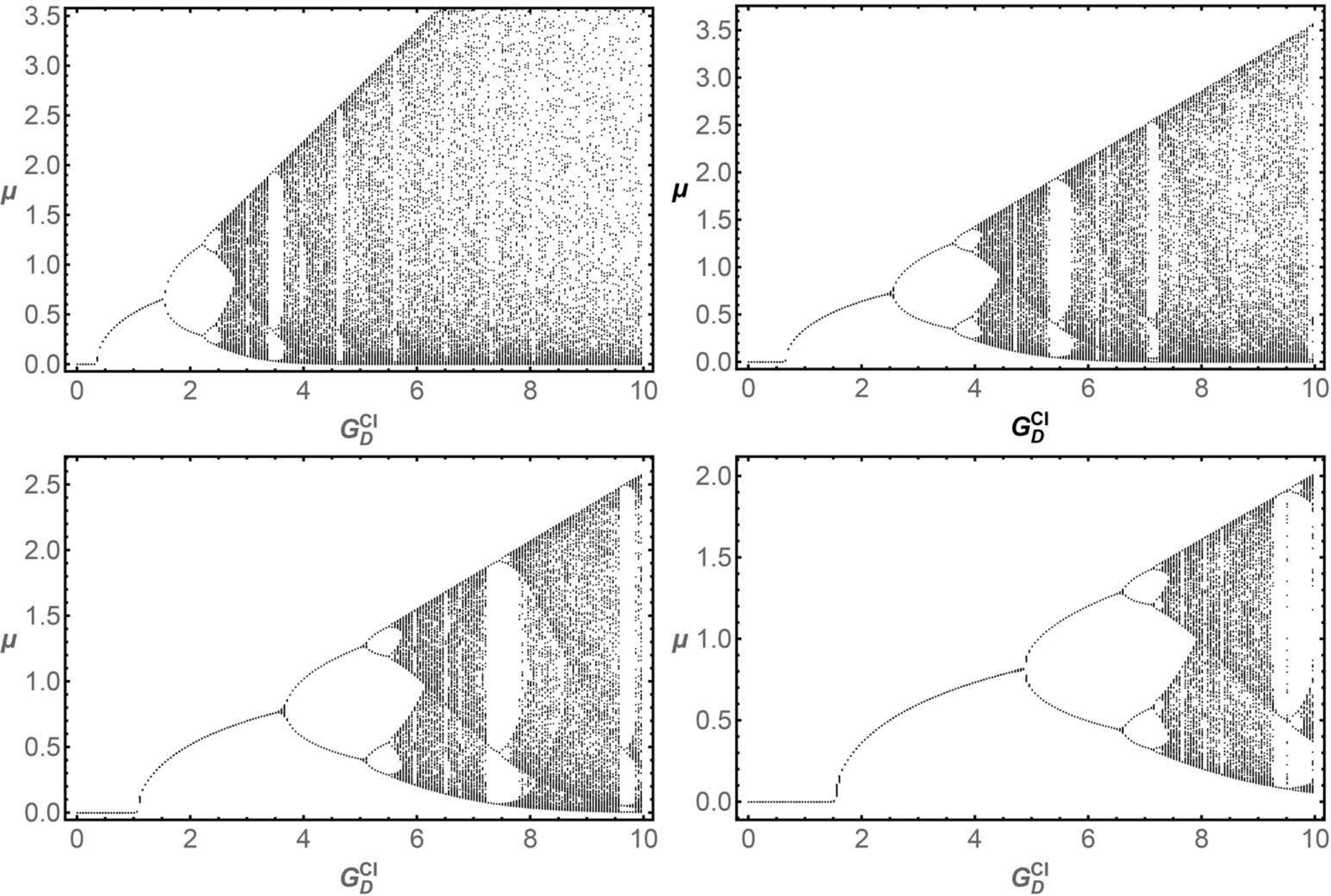}
\caption{Orbits of the gap eq.~(\ref{CI5}) in the chiral limit, with fixed infra-red cut-off $\rho=4.2$  a different  dimensions.  {\em Upper left panel:}  $D=2$;  {\em upper right panel:}  $D=3$;  {\em lower left panel:}  $D=4$; {\em lower right panel:} $D=5$. Procedure is the same as in Fig.~\ref{figNJD4}.}
\label{figCI4}
\end{center}
\end{figure}

\begin{figure}[t!]
\begin{center}
\includegraphics[width=0.9\columnwidth]{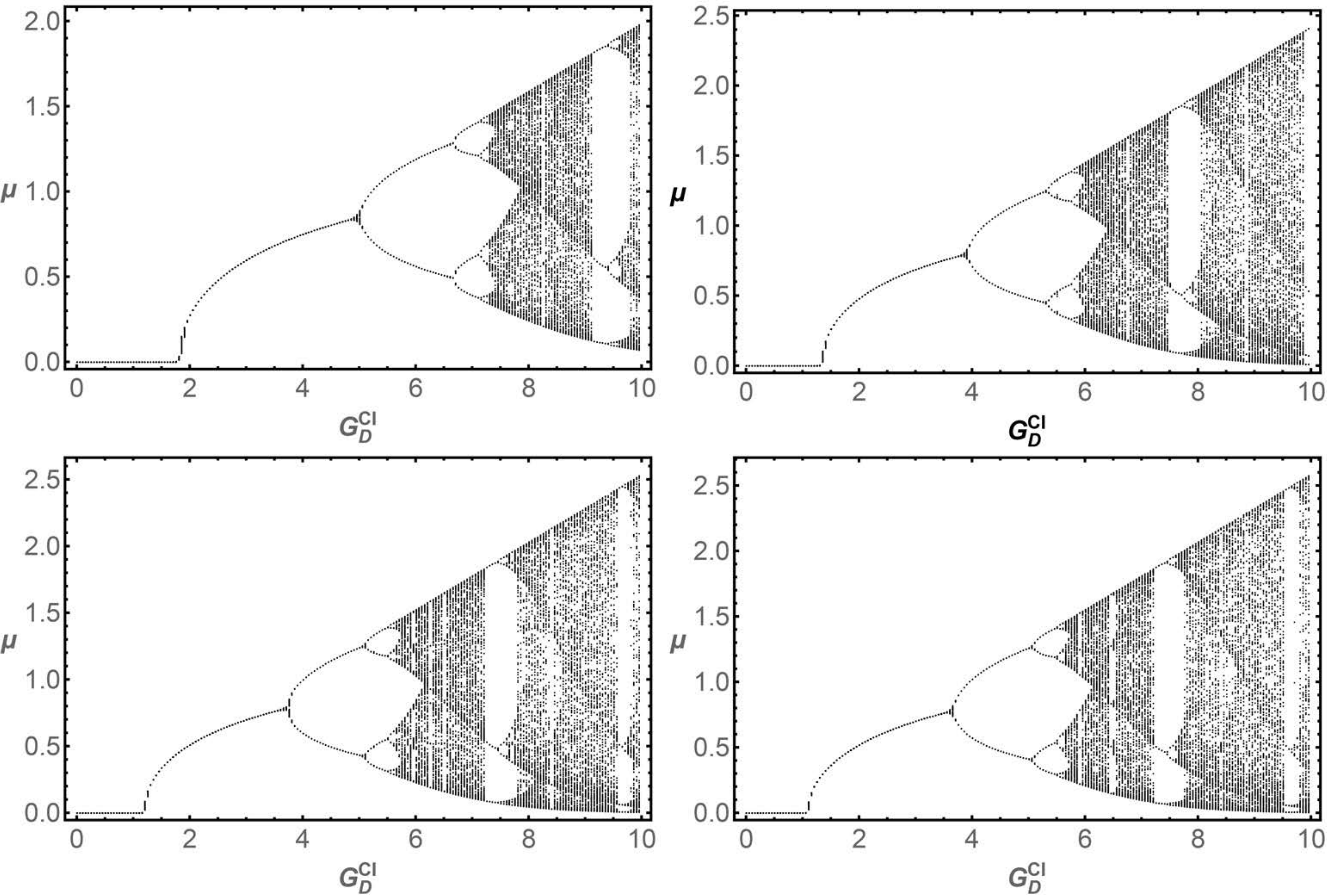}
\caption{Orbits of the gap eq.~(\ref{CI5}) in the chiral limit,
with fixed dimension $D=4$, at different values of the cut-off $\rho$. {\em Upper left panel:}  $\rho=1.2$; {\em upper right panel:} $\rho=1.5$; {\em lower left panel:}  $\rho=2$; {\em lower right panel:}  $\rho=4$. Procedure is the same as in Fig.~\ref{figNJD4}. }
\label{figCI5}
\end{center}
\end{figure}

Finally, we notice the appearance of doubling period orbits for larger values of the coupling. In Fig.~\ref{figCI4} we show this is the case for fixed $\rho$ in different arbitrary dimensions. We observe that the effect of the dimension influences directly the length of the dynamical chiral symmetry branch, between the first and second bifurcations. It also can be seen in the maximum and minimum values of the orbits at fixed $G_D$ in the chaotic domain --bifurcations are wider for low dimensions in the range of couplings plotted. In Fig.~\ref{figCI5} we depict a similar bifurcation pattern for fixed $D=4$ and different  values of $\rho$. In this case we notice that for larger $\rho$, orbits saturate and tend to the same pattern as in the ordinary NJL model shown in Fig.~\ref{figNJD4}. All these observations are consistent with the statement that the chaotic behavior is a robust feature of the  PT regularization of the model, with one or two cut-offs  in arbitrary space-time dimensions.

\section{Summary and Perspectives}\label{conclusions}
By using an iterative procedure for the solution of the gap equation for the NJL model, we explore the behavior of the dynamical quark mass as a function of the coupling strength in two variants of the PT regularization of the model. We naturally find a critical value $G_D^0$ for the coupling  required for breaking chiral symmetry in arbitrary dimensions.  Such critical value specifies the minimal value of the coupling where the dynamical mass bifurcates away from the trivial value $\mu=0$. The dynamical solution is a fixed point and behaves as an attractor for the iterative procedure as expected for an interval of values of $G$. Nevertheless, a second bifurcation occurs when the coupling exceed the value $G_D^*$, where the dynamical solution ceases to be an attractor for the iterations. Larger values of the coupling trigger a cascade of doubling-period orbits~\cite{strogatz} until the gap equation becomes chaotic in the dynamical systems sense. This happens for arbitrary dimensions and even avoiding quark production thresholds in the regularization. 

On physical grounds, external agents like temperature or density dilute the strength of the interaction and induce a dimensional reduction of the system when the temperature and/or chemical potential are very large. A uniform  magnetic field, on the other hand, plays an opposite role and in spite of the fact of inducing a dimensional reduction, it increases the effective strength of the coupling and hence it could drive the effective strength of $G$ to very large values. Therefore, the simple minded exercise carried out in this article allows to address the qualitative behavior of the dynamical mass as a function of the effective coupling in different arbitrary dimensions. The stable branch of the bifurcation has a finite length in this view and thus, the appearance of the second bifurcation sets bounds on the validity of the PT regularization in the sense of stability of the iterative procedure to solve the gap equation.

At this moment, it remains unclear whether the interpretation of every iteration as a self energy insertion in the propagator advocated in~\cite{Aoki:2013gda} is physically reliable or if it is regularization scheme dependent, if the gap equation within the PT (and PV and, perhaps, others) regularization schemes do not admit an iterative solution for super-strong couplings or if the nature of the chaotic behavior found in this article has implications in actual physical systems. Research along these lines is being carried out and results will be presented elsewhere.

\section{Acknowledgments}
AA acknowledges the Department of Physics Gomal University (Pakistan) and IFM-UMSNH, (Mexico) for support. AA also thanks to the staff of IFM-UMSNH, (Mexico) and Paola Rioseco for the hospitality during the visit where part of this work was carried out. AM and AR acknowledge Consejo Nacional de Ciencia y Tecnolog\'ia (Mexico) for support under Grant 256494.

\end{document}